\documentclass[manuscript]{aastex}

\shorttitle{Near IR Adaptive Optics Imaging of NGC 2024}
\shortauthors{Beck, Simon \& Close}

\begin{document}


\title{Near Infrared Adaptive Optics Imaging of the Embedded Cluster NGC 2024}


\author{Tracy L. Beck\altaffilmark{1,2}, M. Simon\altaffilmark{2} and 
 L. M. Close\altaffilmark{3}}


\altaffiltext{1}{Gemini Observatory, 670 N. A'ohoku Pl. Hilo, HI 96720}
\altaffiltext{2}{Department of Physics \& Astronomy, SUNY Stony Brook, Stony Brook, NY, 11794-3800}
\altaffiltext{3}{Steward Observatory, University of Arizona, 933 North Cherry Avenue, Tuscon, AZ 85721-0065}


\begin{abstract}
  
We present the results of a high resolution near infrared adaptive optics survey of the young obscured star forming region NGC 2024.  Out of the total 73 stars detected in the adaptive optics survey of the cluster, we find 3 binaries and one triple.  The resulting companion star fraction, 7$\pm$3\% in the separation range of 0.$''$35-2.$''$3 (145-950 AU), is consistent with that expected from the multiplicity of mature solar-type stars in the local neighborhood.  Our survey was sensitive to faint secondaries but no companions with $\Delta$K$'$ $>$ 1.2 magnitudes are detected within 2$''$ of any star.  The cluster has a K$'$ luminosity function that peaks at $\sim$12, and although our completeness limit was 17.7 magnitude at K$'$, the faintest star we detect had a K$'$ magnitude of 16.62.

\end{abstract}




\section{Introduction}

Recent observational studies have shown that the typical result of star formation is a binary or higher order multiple star system.  Duquennoy \& Mayor (1991) found that more than half of the nearby, mature solar-type stars have stellar companions.  Studies of the young Trapezium star forming region in Orion (age $\sim$0.5 My) have found a multiplicity consistent with the local neighborhood (McCaughrean \& Stauffer 1994, Prosser et al. 1994, Petr et al. 1998, Simon, Close \& Beck 1999). Rigorous observational investigations toward stars in the nearby young dark cloud complexes (Taurus-Auriga, Ophiuchus, Chameleon, Lupus, Corona Australis; ages range from 1 to a few My) reveal that the binary frequency in these regions is significantly higher (Ghez, Neugebauer \& Matthews 1993; Leinert et al. 1993; Reipurth \& Zinnecker 1993; Simon et al. 1995; Ghez et al. 1997; K\"ohler \& Leinert 1998).  Studies of older regions, such as the Pleiades and $\alpha$ Per (Bouvier, Rigaut \& Nadeau 1997, Patience et al. 2002) find multiplicities consistent with the solar value.  For the Hyades cluster (660 My), Patience et al. (1998) found a multiplicity that was intermediate between the local solar neighborhood and the young stars in the dark cloud regions, which in turn led to their speculation that the binarity of a region may evolve in time.  Prosser et al. (1994) and Bouvier, Rigaut \& Nadeau (1997) proposed that the multiplicity may depend on the star formation efficiency and stellar density of a region.  

Comparison of the individual multiplicity studies is difficult because each surveys a different range of companion brightnesses and separations.  Duch\^{e}ne (1999) devised a method which takes dynamic range, sample completeness and the chance projection of background objects into account and reanalyzed the available data from past multiplicity surveys.  His overall results were similar to previous findings; the Trapezium, Pleiades, Hyades and optical studies of additional sites of star formation in Orion show essentially the same multiplicity as expected from solar-type stars in the local neighborhood, and the dark cloud complexes have a higher multiplicity at a 3$\sigma$ confidence level.
 
The results presented by Patience \& Duch\^{e}ne (2001) suggest that binarity may be more dependent upon formation environment rather than cluster age.  At the present time, however, it is still uncertain how the properties of a parent molecular cloud affect the duplicity of stars.  Hence, to study the multiplicity of young stars in diverse environments and further test these ideas, we have made a high resolution adaptive optics K$'$-band survey of NGC 2024 in Orion.  NGC 2024 is an important target for this study because it is comparable age to the Trapezium region ($\sim$ 0.3 My; Meyer 1996) but has a stellar density and star formation efficiency (SFE) intermediate between the dense clusters and the sparse dark cloud complexes (Lada et al. 1991).  If the multiplicity of a region depends upon these parameters then one might expect NGC 2024 to be intermediate in this regard also, larger than the low multiplicity Trapezium, yet smaller than the dark clouds where multiple star systems dominate.

NGC 2024 is a spectacular H II emission region in the Orion (Lynds 1630) molecular cloud, and is about 15$'$ to the east of $\zeta$ Ori at a distance of $\sim$415 pc. (Anthony-Twarong 1982; Brown, de Geus \& de Zeeuw 1994).  It also contains an embedded cluster of young stars (Grasdalen 1974).  The central region of the cluster is obscured by a dark lane; a protrusion of dense material from the background cloud results in a significant visual extinction (A$_{v}$ $\sim$10-20 magnitudes, Barnes 1989; Meyer 1996) toward the cluster core.  Inspection of Digitized Sky Survey images and archival HST data reveal no stars detectable in visible light toward the core of the cluster.   For this reason, most of the infrared sources in NGC 2024 are likely cluster members.  Furthermore, contamination by background stars is small because of the region's large column density of background molecular material and high latitude (b $\sim$ 16$^{\circ}$) toward the galactic anti-center.  In section 2 we discuss the details of the observations, the method of data reduction, and the sensitivity and results of the AO corrected data.  We present and the results of the multiplicity survey in \S3 and discuss briefly the stellar population of the NGC 2024 cluster in \S4.  

\section{Observations}

The observations for this project were made on 3 and 4 November 1998 at the Canada-France-Hawaii 3.6 meter telescope on Mauna Kea.  We used the facility AO system PUEO, the Adaptive Optics Bonnette, which is a 19 element curvature sensing system (Rigaut et al. 1994) with KIR, the AO camera which contains a 1024 $\times$ 1024 HAWAII detector (Doyon et al. 1998).  We used a K$'$ filter and a plate scale of 0.$''$0348 per pixel.   A total of 12 fields, each 35.$''$6 on a side, were observed with the AO system using NGC 2024 IRS 1 (R $\sim$ 11 mag; Star AO 33 in Table 1) and AO 25 (IRS 12 from Barnes et al. 1989; R $\sim$ 15 mag) as reference stars.  Four frames of the cluster core were observed without adaptive optics, as suitable reference stars could not be identified within $\sim$45$''$ of the target region because of the very large obscuration at wavelengths required for wavefront correction.

Four observations in a 5$''$ square dithered pattern were made for each of the AO fields at integration times of 2 seconds and 60 seconds.  The images were flat-fielded and the median value of each pixel was determined to produce a frame used for sky subtraction.  To create long and short exposure frames at each field position, the images were registered, shifted, and combined using reduction procedures in IRAF.  A point spread function (PSF) was generated from the single stars in each frame.  Photometry was done both with a 1.$''$6 circular aperture and by fitting the PSF to all of the stars using the DAOPHOT routines. For single stars, the results from these two methods were always consistent to within two tenths of a magnitude.  Bright stars (K$'$ $<$ $\sim$8) were saturated in the long exposure image, so their magnitudes were determined by comparison with fainter stars that were detected in both the long and short exposure frames.  

The conditions were not photometric during our observations at the CFHT, so we were not able to flux calibrate the images.  We could, nevertheless, derive relative fluxes for all of the stars by using sources in common in over-lapping images.  Subsequently, we calibrated the stars with AO numbers 28, 46, 48, 51 and 57 in Table 1 by imaging in the K$'$-band on two occasions; using NSFCam at the NASA-IRTF on 02 January 2002 and using NIRC2 at the Keck Observatory on 13 March 2002.  The derived K$'$ magnitudes of these calibration stars were consistent to within $\pm$0.05 mag.  Using the K-K$'$ to H-K relation from Wainscoat \& Cowie (1992),  K$'$-K=(0.18-0.22)H-K, we also estimated the K$'$ magnitudes of these stars based on previously reported simultaneous measurements of the H and K magnitudes and found no evidence for variability in these sources.  We estimate the accuracy of our photometry to be $\pm\sim$0.08 magnitudes.

Figure 1 shows the final mosaic of the 16 data frames; its total area is $\sim$4.8 square arc minutes.   Figure 2 plots the detected stars in pixel (x,y) coordinates (see also Table 1).   Approximately 3.7 square arcminutes of the mosaic image is AO corrected (see Figure 1).  On these images, the full-width at half-maximum (FWHM) of the wavefront reference star PSF was 0.$''$17, very close to the diffraction limit of the CFHT in the K$'$-band.  The uncorrected images had average FWHM values of $\sim$0.$''$9.  The AO guide stars (AO 33 and 25) had Strehl ratios of 23\% and  and 12\%, respectively; perfect adaptive optics correction yields Strehls of 100\%.  The Strehl ratio of AO 86, a star in a frame obtained with the AO off, was $\sim$1.7\%.  Figures 3a and 3b present surface plots of AO 33 and 86, normalized to the same flux, and demonstrating the vastly improved angular resolution of the adaptive optics corrected images.  

Table 1 presents the 94 stars detected in this survey as the AO star number, x and y positions on the mosaic, RA, Dec and the K$'$ magnitude from this survey.  The RA and Dec coordinates were derived by precessing the 1950 coordinates from Meyer (1996) to epoch 2000 and were doublechecked for accuracy by pixel offsets from the astrometric position of NGC 2024 IRS 1 (AO 33).  The isoplanatic patch for these observations was $\sim$35$''$; the 5$\sigma$ completeness limit within a 35$''$ radius of the AO 33 reference star is $\sim$18.6, and is $\sim$18.2 for frames obtained using AO 25 for correction.  We estimate our overall completeness to be K$'$=17.7 mag, this is the 5$\sigma$ detection limit at the outer edge of the AO survey, $\sim$55$''$ from AO 25.   Figure 4 shows the distribution of K$'$ magnitudes of the 94 stars we detected toward NGC 2024.   

In the deep exposure adaptive optics frames, the bright stars (K$'$ $<$ 9) all have associated ghost images at a position angle of $\sim$300$^{\circ}$ and separation of $\sim$11$''$.  These ghosts are out of focus back reflections off the PUEO dichroic and are of much lower Strehl than the point sources in the AO field.  Since the ghosts were easily identifiable, they were all removed from the mosaic image (Figure 1).  Overplotted in Figure 4 is the distribution of the integrated magnitude of the ghosts associated with the bright stars in the field.  We have photometered the ghost images and found a signal-to-noise $>$ 3 on four that have K$'$ magnitudes $>$ 17.7.  The faintest star we detect toward the cluster has a K$'$ magnitude of 16.62 (AO 41) and signal-to-noise of $\sim$25, and the faintest ghost has a total K$'$ magnitude of 18.6 and an integrated signal-to-noise of $\sim$4. 




\section{The Multiplicity of NGC 2024}

Of the 94 stars detected in the total survey sample, we find 78 single stars, 5 binaries and 2 triples in the separation range 0.$''$18 to 2.$''$3 (75-950 AU).  The lower limit is defined by the closest system detected, and the upper bound corresponds to the limit adopted for the survey of Padgett, Strom \& Ghez (1997).  Figures 5 a-g show contour plots of these seven multiple systems, and Table 2 lists the AO primary star number, AO companion star number, K$'$ magnitudes of the primary, $\Delta$K$'$ of the companion and its projected angular separation.  AO 17/18 is the closest binary in the survey with a projected separation of only 0.$''$18.  Figure 5e presents four stars which lie within $\sim$5$''$ of each other.  AO 69, however, is probably not associated with the other three because it is further away (2.$''$9 away from AO 67) and it is redder based on the J, H and K-band photometry of Meyer (1996).  The faintest binary star system that we detect is AO 63/64, at magnitudes of 13.32 and 13.38, respectively.   Figure 5f and 5g present systems detected in frames without AO correction.  

Combining the non-AO sample with the adaptive optics data results in a lower sensitivities to faint companions.  The limiting magnitude for the detection of single stars in the non-AO frames is $\sim$14.4 mag.  Thus, in order to ensure completeness to companions with a mass ratio down to q$\sim$0.1 in the non-adaptive optics frames ($\Delta$K$'$=3.3 magnitudes), only single stars brighter than 11.1 can be considered for the statistics.  Subsequently, if we combined the AO and non-AO frames and analyzed them for multiplicity in different separation ranges, we would only be able to consider stars with K$'$-band magnitudes greater than 11.1 for mass ratio completeness.  This introduces unnecessarily low sensitivity limits on the adaptive optics images.  Hence, for statistical analysis of the multiplicity, we consider only the 73 stars detected in the frames observed with adaptive optics correction.  This restriction excludes the binary and triple system shown in Figure 5f and 5g.

To estimate the number of multiples which could be chance projections through the cluster, we analyze the surface density as a function of stellar separation using the method of Simon (1997; Figure 6).  At large separations, the surface density $\Sigma(\theta)$ versus separation $\theta$ represents the large-scale density distribution of the underlying cluster.  On the smaller scale, the break in the $\Sigma(\theta)$ curve is significant at the $\sim$2.5 sigma level and represents the separation in which binaries become an important contribution to the surface density of stars (Larson 1995; Simon 1997).   For the 73 stars detected in the AO frames, Figure 6 shows that the break occurs at $\sim$2$''$, a projected linear separation of $\sim$830 AU,  in agreement with the 2.$''$3 upper bound adopted for the AO data.  For separations greater than this value, the distribution of stars appears random with a roughly constant surface density, $\Sigma_{ave}$ $\sim$ 7.7$\times$10$^{-4}$ stars/sq. arcsec.

It is important to determine the sensitivity to companion brightness and separation in order to assess the completeness of the survey.  In practice, the sensitivity to companions decreases with decreasing angular separation from a central star.  Additionally, the sensitivity to NGC 2024 members decreases with increasing distance from the adaptive optics wavefront reference star because of anisoplanatic effects.  In Figure 7 we present the sensitivity ($\Delta$K$'$) versus angular separation for three stars  at varying distances from the AO guide star.  The sensitivity at the projected separation from the central star is essentially the limiting magnitude; five times the standard deviation over the mean counts in each 0.$''$13 wide (4 pixel) annuli.  Naturally, $\Delta$K$'$ depends on the magnitude of the star for which it is determined.  For this reason, we have also included in Figure 7 the sensitivity curve for a faint (K$'$ = 15.60) star.  This star was 53$''$ from the wavefront reference star; the sensitivity to companions at large separations ($>$ 3$''$) around this star confirms our overall magnitude completeness limit of K$'$=17.7 (see Figure 7).  Over-plotted as triangles in the figure is the $\Delta$K$'$ versus angular separation for all possible primary-companion permutations with separations less than 3.$''$5 detected in the survey (i.e., the ``quadruple star'' presented in Figure 5e appears as four different primary-companion systems).

We tested further the accuracy of the sensitivity curves in Figure 7 in two ways:  

1) We subtracted an average PSF from a single star and added five times the average of the absolute value residual counts in each of the annuli.  This gave a second estimate of the limiting magnitude.  The sensitivity at each separation using this method was typically fainter by a few tenths of a magnitude than that presented in Figure 6.

2)  We constructed binary models based on the $\Delta$K$'$ and companion separations presented in Figure 7.  An average PSF was subtracted from the synthesized binary at the position of the primary in the hope of detecting the companion in the model.  An example of one of these tests is presented in Figures 8a and b.  The model companion, 4 magnitudes fainter than its primary at a separation of 0.$''$6, is easily detected in the residual image (Figure 8b).

We believe that the sensitivity curves presented in Figure 7 are conservative estimates.  For separations greater than $\sim$1$''$, companions brighter than the completeness limit of 17.7 magnitudes should have been detected for all stars with K$'>$9.  No companions with $\Delta$K$'>$1.2 mag are found for systems with separations less than 2$''$.  Because of the obvious lack of faint companions at small separations, corrections for incompleteness seem unnecessary.  However, for consistency with the study of Duch\^{e}ne (1999), we redefine the lower limit for binary separations to be 0.$''$35.  This is the smallest separation which ensures a sensitivity to companions with a mass ratio of 0.1, assuming a $\Delta$K magnitude of $\sim$3 (Duch\^{e}ne 1999).

Given the average surface density of stars, $\Sigma_{ave}$, there is a 100\% chance of finding a star within 20.$''$3 of another.  Thus, in the 0.$''$35 - 2.3$''$ separation range in which we consider stars to be associated, there is a 1.3\% probability for stars which are not physically associated to be mistaken for a multiple system.  Out of the 73 stars detected in the AO corrected frames, we find three binaries and one triple in this separation range (Figure 5a, c, d and e) and for statistical purposes one of these must be considered a chance projection.  For this case, the companion star fraction (csf), the total number of companions per primary star, is defined as:

\begin{eqnarray}
csf & = & \frac{B+2T-C}{S+B+T}
\end{eqnarray}

Where B is the number of binaries, T is the number of triples, S is the number of singles and C is the number of companions which are likely chance projections.    The csf for NGC 2024 is (3+2-1)/(54+3+1) = 7$\pm$3\%.  The number of single stars used in our statistical analysis excludes 6 bright sources that are likely more massive than 1 solar mass.  We also exclude five single stars that are too faint for detection of a 0.1 mass ratio companion in the separation range we define.  The uncertainty in the csf is derived from Poisson statistics. 

We follow the statistical analysis described by Duch\^{e}ne (1999) to compare the observed multiple systems in the NGC 2024 AO fields to the multiplicity of solar-type stars in the local neighborhood.  The 0.$''$35 - 2.$''$3 limits imposed on the AO data correspond to a projected separation range of 145-950 AU at the 415 pc distance of NGC 2024.  Using corresponding orbital periods as limits, we integrate the period distribution presented by Duquennoy \& Mayor (1991) and find that out of 54 stars, $\sim$3 companions would be expected from the field distribution.  This corresponds to a companion star fraction of 6$\pm$2\%, consistent with our results.  Hence, we find no evidence for an excess multiplicity toward NGC 2024.  Levine et al. (2000; 2002) also surveyed the central region of NGC 2024 at high spatial resolution and report an enhanced multiplicity with respect to sun-like stars in the solar neighborhood.  A comparison of their list of detected objects, when they are published, with ours may explain the difference between the results of the two surveys.

Padgett, Strom \& Ghez (1997) observed NGC 2024 using the F814W (I-band) filter with WFPC2 on the Hubble Space Telescope and detected 8 binary stars in the separation range 0.$''$3 - 2.$''$3.  In their paper, they combine the results of the NGC 2024 study with the young star forming regions NGC 2068 and 2071 in Orion.  The sample size for the NGC 2024 region for their study was kindly shared by D. Padgett (private communicaton).  Based on this sample size and using the analysis method of Duch\^{e}ne (1999), we estimate a companion star fraction of 11$\pm$4.0\% for NGC 2024 by excluding two companions detected by Padgett, Strom \& Ghez (1997) in the statistics; one companion is fainter than the q$\sim$0.1 mass limit and 1 is likely a chance projection.  Because of the large uncertainties from the small number statistics involved, the difference in multiplicity between this value and the one we derive for our AO survey is less than 1$\sigma$.  

Of the 8 binaries identified by Padgett, Strom \& Ghez (1997), we clearly detect four (AO 16/20, AO 23/24, AO 17/18 and AO 54/55) and another two are out of the field of view of our survey.  We do not detect two companions, faint counterparts to AO 8 and AO 5, that are discovered at a 5 and 2.5$\sigma$ confidence level in their study.  If we apply a F814W to K$'$ correction based on the magnitudes of the primaries, both companions would have K$'$ magnitudes of $\sim$17.5.  These stars are closer than 0.$''$45 from their respective primaries and were fainter than the detection limit of our AO survey at these separations.

Figure 9 plots the companion star fraction, normalized by the csf value for main sequence stars in a comparable sensitivity sample, versus the density of stars for seven regions of star formation.  The companion star fraction for NGC 2024 is from this study, and the remaining csf data is taken from Duch\^{e}ne (1999).  The references for densities of the young star forming regions are: Taurus- Gomez et al.(1993), Chamaeleon - Gomez \& Kenyon (2001), Lupus- Nakajima et al. (2000), Ophiuchus and IC 348 - Luhman et al. (2000), NGC 2024 - Lada et al. (1991), Orion Trapezium -  Luhman et al. (2000).  We have excluded Corona Australis (Wilking et al. 1997) from this plot as the multiplicity of this star forming region is more uncertain because of small number statistics (csf/csf$_{MS}$ value of 3$\pm$2).  Although the uncertainties in the individual datapoints are significant, Figure 9 suggests a possible inverse correlation between the stellar density of a region and its efficiency at forming multiple star systems, as previously discussed in the works of Prosser et al. (1994) and Bouvier, Rigaut \& Nadeau (1997).   The figure also indicates that establishing a relationship between companion star fraction and the density of a star forming region at a reasonable level of significance will require sample sizes much larger than those achieved thus far.





\section{The Stellar Population of NGC 2024}

AO 33 and AO 92 are the brightest infrared sources toward NGC 2024, they are the NGC 2024 IRS \# 1 and \# 2 sources of Grasdalen (1974), respectively.  A turnover at K $\sim$ 12 mag was first discovered toward NGC 2024 by Lada et al. (1991).  Hodapp (1994) and Meyer (1996) also found a similar magnitude distribution.   In Figure 4, the K$'$ brightness distribution of the detected sources peaks at $\sim$12 mag.   Out of the total 94 stars detected in the survey, only 4 are fainter than $\sim$15 magnitude.  Our detection of ghost images down to (and beyond) the estimated magnitude completeness suggests that there is not a large stellar or sub-stellar population of NGC 2024 members in the range of 16$<$K$'$$<$18 that was missed by our survey.  

It is not possible to characterize the nature of the objects we have detected in detail because our observations were made in only one photometric band.  In the absence of color information, we cannot distinguish between a population of reddened low mass stars and a population of brown dwarfs.  We note, however, that according to Baraffe et al.'s (1998) models, a 1 million year old star at the substellar boundary should have an unobscured K-band magnitude of $\sim$13 at the distance of NGC 2024.   A star at the substellar boundary, observed through the average extinction of $\sim$2 mag at K (Meyer 1996), would therefore appear to be $\sim$15 mag in our survey.  However, out of the total 94 stars detected in our study, we find only four that are fainter than K$'$= 15 mag.  This suggests that a significant population of brown dwarfs, comparable to those recently discovered toward the Trapezium (Luhman et al. 2000; Lucas et al. 2000; 2001) has eluded our survey of the cluster core.   

Comer\'{o}n, Reike \& Reike (1996) studied the low mass population of NGC 2024 and found numerous sources in the region that have K-band magnitudes greater than $\sim$15.  They classify many of these as brown dwarf candidates down to 0.04 solar masses, based on the sub-stellar models of Burrows et al. (1993).   Levine et al. (2002) have made a similar sensitivity K-band survey of the central 30 square arcminutes of NGC 2024 and have also identified what they believe is a significant population of brown dwarfs in the region.  A population of brown dwarfs may fall below the sensitivity limits of our survey because of low mass or large obscuration, or be located outside the region we have observed.   If BDs are stellar embryos that were ejected during the formation of a multiple system, as suggested by Reipurth \& Clarke (2001), in a region as young as NGC 2024 such a population may observable as an overdensity of brown dwarfs in the periphery of the cluster.  Hence, we believe NGC 2024 is a good candidate for follow-up observations to test for gradients in the detectability of brown dwarfs in various regions in the cluster.  

\section{Summary}

We observed a total area of 4.8 square arcminutes toward the young embedded cluster NGC 2024 and detected 94 stars in total, 73 in frames observed with adaptive optics correction.  In the separation range of 0.$''$35 to 2.$''$3 for the AO frames, we find 3 binaries and 1 triple for a companion star fraction of 7$\pm$3\%.  The multiplicity of NGC 2024 is not significantly different from that expected from the solar-type stars in the local field.  Moreover, we find a turn-over in the magnitude distribution of stars at a magnitude of $\sim$12, and comparably few faint members.





\acknowledgments

We would like to thank the staff and the PUEO team at the CFHT for making the observations for this project possible.  We also thank to Joanna Levine for discussion of her survey and work in progress, Deborah Padgett for sharing information on her 1997 study and the referee for a helpful report.  The work of TB and MS was supported through NSF Grant 98-19694.  This study was supported in part by the Gemini Observatory, which is operated by the Association of Universities for Research in Astronomy, Inc., on behalf of the international Gemini partnership of Argentina, Australia, Brazil, Canada, Chile, the United Kingdom, and the United States of America.

\clearpage


\figcaption{ The near infrared (K$'$-band) mosaic of NGC 2024 constructed from 15 individual 1024$\times$1024 array images of the region.  The stars to the right of the line are from frames that were obtained with AO correction, and stars to the left were observed without AO.  North is up, East is to the left and the 0, 0 position of the image corresponds to an RA(2000) and Dec(2000) of 05 41 37.7 -01 54 32, which is the position of NGC 2024 IRS \#1 (AO 33 in Table 1). }

\figcaption{ A diagram of the x and y positions for all of the stars detected in the adaptive optics survey of NGC 2024 (see also Table 1).  The 0, 0 position in this figure corresponds to an RA(2000) and Dec(2000) of 05 41 46.2 -01 55 23 on the sky. }

\figcaption{Fig. 3a is A surface plot of AO 33, the wavefront reference star, from a frame with the AO on.  The strehl ratio is $\sim$23\%, and the FWHM = $\sim$0.$''$17.  Fig. 3b is A surface plot of AO 86 from observations made with the AO off.  The star has been normalized to the same flux as AO 33 in Figure 3a and the plot is scaled to the same z range.  The strehl ratio of the star is $\sim$1.7\%, and the FWHM = $\sim$0.9$''$.}

\figcaption{ The distribution of K$'$ magnitudes of the total 94 stars detected in the survey.  Overplotted is the integrated magnitudes of the ``ghost'' images associated with each bright star, and thus suggests that a significant population of faint stars was not missed by our survey.}

\figcaption{ Contour plots of the multiple systems with projected separations less than $\sim$2.$''$3 detected toward NGC 2024.  In all of the frames, North is down and East is to the left.  Out of 94 stars in the total sample, there are 5 binaries and 2 triples.  For all of the plots except e), the contour levels are 14,28,42,56,70,84 and 96\% of the peak.  For e) the contour levels are 5,8,11,14,28,40,60,90\% of the peak to increase the dynamic range, the star in the lower left corner of this frame is likely not associated with the other three because it is further away and redder in H-K color (as determined by photometry from Meyer 1996).  The closest separation companion is b) at 0.$''$18 (it was $\sim$40$''$ away from the wavefront reference star). f) and g) were in frames where the AO was turned off.}

\figcaption{ The Surface Density plotted versus separation for the 73 stars detected in the AO fields.  The break in the curve occurs at a $\theta$ of $\sim$2$''$ and represents the separation range where binaries become an important contribution to the surface density of stars.}

\figcaption{ The senstivity, $\Delta$K$'_{inst}$, versus separation, for four stars with K$'$=12.14, 12.15, 12.22 and 15.60.  The number in parenthesis in the upper left is the angular distance from the AO wavefront reference star in seconds of arc.  The triangles represent all of the systems with separations less than 3.$''$5 detected in the survey; the magnitude of the brighter star is overplotted.}

\figcaption {Fig. 8a presents the residual emission around a 12.1 magnitude star after a 0.$''$9 radius PSF was subtracted.  The field of view is 2$''\times$2$''$ and the apparent ``disk-like'' emission is an artifact of the AO corrected image.  To demonstrate the residual structure, this image is displayed from -100\% to 100\% of the peak flux.  Overplotted is a contour plot of the image residuals, contour levels are 30, 45, 60, 75 and 90\% of the peak.  Fig. 8b - The same 12.1 magnitude star after subtraction of a 0.$''$9 radius PSF but a ``companion'' has been added to test the derived binary sensitivity curve.    Fig. 8b is a contour plot of the residual emission, again the field of view is $''\times$2$''$ and the contour levels are 30, 45, 60, 75 and 90\% of the peak.  The companion has a $\Delta$ K$'$ of 4 magnitudes and a separation of $\sim$0.$''$6.  The position angle was chosen to correspond with the peaks of the emission after model subtraction (observed in Figure 8a). The faint companion is easily detectable in the image.}

\figcaption{ The companion star fraction, normalized by the value for the solar neighborhood population, plotted versus the density of stars in the region.}






\clearpage

\begin{deluxetable}{ccccccc}
\tabletypesize{\scriptsize}
\tablecaption {Detected Stars in NGC 2024. \label{tbl-1}}
\tablewidth{0pt}
\tablehead{
\colhead{AO} & \colhead{Name$^{a}$} & \colhead{x position}  & \colhead{y position}  &
\colhead{RA J(2000)}  & \colhead{Dec J(2000)} & 
\colhead{K$'$ mag} 
}
\startdata
 1 &  &  5020 &  3495  & 05 41  34.9 & -01 53  23 &  12.61  \\
 2 & &   4860 &  2030 &  05 41  35.3 & -01 54  14  &  13.22  \\
 3 & &   4720 &  2920  &  05 41  35.6 & -01 53  43  &  14.37  \\
 4 &  &  4710 &  1960 &  05 41  35.7 & -01 54  17 &  13.84  \\
 5 &  &  4645 &  3240  &  05 41  35.8 & -01 53  32 &  13.38  \\
 6 & IRS 18 &  4540 &  1895 &  05 41  36.1 & -01 54  19  &  8.62  \\
 7 &  &  4475 &  3600 &  05 41  36.2 & -01 53  19  &  13.32  \\
 8 &  &  4465 &  3175 &  05 41  36.2 & -01 53  34  &  11.68  \\
 9 &  &  4430 &  2930 &  05 41  36.2 & -01 53  42  &  15.69  \\
 10 &  &  4420 &  4025 &  05 41  36.3 & -01 53  4  &  12.95  \\
 11 &  &  4440 &  1390 &  05 41  36.3 & -01 54  36  &  12.74 \\
 12 &  &  4370 &  3760 &  05 41  36.4 & -01 53  14  &  11.90  \\
 13 & IRS 17 &  4360 &  2735 &  05 41  36.5 & -01 53  49  &  9.78  \\
 14 &  &  4335 &  2150  &  05 41  36.5 & -01 54  10  &  14.20  \\
 15 &  &  4325 &  2340 &  05 41  36.5 & -01 54  3  &  11.10  \\
 16 &  &  4260 &  4750 &  05 41  36.7 & -01 52  39  &  11.49  \\
 17 & IRS 16 &  4260 &  2620 &  05 41  36.7 & -01 53  54  &  11.13  \\
 18 & IRS 16 &  4260 &  2615  &  05 41  36.7 & -01 53  54  &  10.63  \\
 19 &  &  4295 &  1210 &  05 41  36.7 & -01 54  43  &  11.81  \\
 20 &  &  4230 &  4710 &  05 41  36.7 & -01 52  41  &  12.68  \\
 21 &  &  4210 &  3160  &  05 41  36.8 & -01 53  34  &  12.62  \\
 22 &  &  4185 &  4910 &  05 41  36.9 & -01 52  35  &  12.15  \\
 23 &  &  4150 &  5040 &  05 41  36.9 & -01 52  29  &  12.18  \\
 24 &  &  4145 &  5030 &  05 41  37.0 & -01 52  30  &  11.62  \\
 25 & IRS 12 &  4015 &  3910  &  05 41  37.2 & -01 53  8  &  8.40  \\
 26 &  &  3950 &  4710  &  05 41  37.4 & -01 52  41  &  11.05  \\
 27 &  &  3910 &  5080  &  05 41  37.5 & -01 52  27  &  13.97  \\
 28 &  &  3930 &  1860  &  05 41  37.5 & -01 54  19  &  10.61  \\
 29 &  &  3920 &  1520  &  05 41  37.5 & -01 54  33  &  11.50  \\
 30 & IRS 15 &  3880 &  2810 &  05 41  37.6 & -01 53  46  &  8.55  \\
 31 &  &  3920 &   570 &  05 41  37.6 & -01 55  5  &  13.69  \\
 32 &  &  3900 &  1380 &  05 41  37.6 & -01 54  36  &  10.73  \\
 33 &  NGC 2024 IRS \#1 &  3845 &  1520  &  05 41  37.7 & -01 54  32  &  5.93  \\
 34 &  &  3845 &  1670  &  05 41  37.7 & -01 54  27  &  9.69  \\
 35 &  &  3760 &  3950 &  05 41  37.8 & -01 53  6  &  9.95  \\
 36 &  &  3750 &  3750  &  05 41  37.9 & -01 53  13  &  12.84  \\
 37 &  &  3730 &  2640  &  05 41  38.0 & -01 53  52  &  10.79  \\
 38 &  &  3740 &  1605  &  05 41  37.9 & -01 54  28  &  11.94  \\
 39 &  &  3675 &  3545  &  05 41  38.0 & -01 53  20 &  14.59  \\
 40 &  &  3720 &  1000  &  05 41  38.0 & -01 54  50  &  12.14  \\
 41 &  &  3740 &   260  &  05 41  38.0 & -01 55  16  &  16.62  \\
 42 &  &  3675 &  2065  &  05 41  38.1 & -01 54  12  &  14.27  \\
 43 & IRS 11 &  3620 &  4015  &  05 41  38.1 & -01 53  4  &  9.01  \\
 44 &  &  3600 &  3460  &  05 41  38.2 & -01 53  23 &  13.66  \\
 45 &  &  3600 &  3320  &  05 41  38.2 & -01 53  28  &  12.70  \\
 46 &  &  3580 &  1840  &  05 41  38.3 & -01 54  20  &  11.26  \\
 47 & IRS 10 &  3480 &  3620  &  05 41  38.5 & -01 53  18  &  8.35  \\
 48 &  &  3460 &  1720  &  05 41  38.6 & -01 54  24  &  13.08  \\
 49 &  &  3340 &  8150  &  05 41  38.9 & -01 54  56  &  13.32  \\
 50 & IRS 14 &  3290 &  2610  &  05 41  39.0 & -01 53  53  &  8.91  \\
 51 & IRS 13 &  3260 &  2150  &  05 41  39.1 & -01 54  9  &  9.73  \\
 52 &  &  3230 &  2490  &  05 41  39.1 & -01 53  57  &  13.27  \\
 53 &  &  3205 &  4060  &  05 41  39.1 & -01 53  3  &  13.29  \\
 54 &  &  3105 &  3485  &  05 41  39.4 & -01 53  21  &  10.71  \\
 55 &  &  3110 &  3500  &  05 41  39.4 & -01 53  22 &  10.24 \\
 56 &  &  3110 &  3290  &  05 41  39.4 & -01 53  28  &  11.41 \\
 57 &  &  3110 &  1360  &  05 41  39.4 & -01 54  37 &  12.69  \\
 58 &  &  2865 &  3240  &  05 41  40.0 & -01 53  30  &  11.69  \\
 59 &  &  2830 &  2660  &  05 41  40.0 & -01 53  51  &  15.60  \\
 60 &  &  2815 &  2330  &  05 41  40.1 & -01 54  2  &  11.89  \\
 61 &   & 2625 &  2875  &  05 41  40.5 & -01 53  43   &  12.97  \\
 62 &  &  2510 &  2890  &  05 41  40.8 & -01 54  3  &  13.83  \\
 63 &  &  2340 &  1445  &  05 41  41.2 & -01 54  33  &  13.38  \\
 64 &  &  2330 &  1450  &  05 41  41.2 & -01 54  33  &  13.32  \\
 65 &  &  2330 &  1250  &  05 41  41.3 & -01 54  39  &  11.63  \\
 66 &  &  2300 &  2745  &  05 41  41.3 & -01 53  47  &  12.22  \\
 67 &  &  2290 &  1405  &  05 41  41.3 & -01 54  34  &  11.33  \\
 68 &  &  2265 &  2600 &  05 41  41.3 & -01 53  53  &  15.98  \\
 69 &  &  2215 &  1370  &  05 41  41.4 & -01 53  34  &  13.76  \\
 70 & &   2205 &  2170  &  05 41  41.5 & -01 54  7  &  12.19  \\
 71 & &   2170 &  2130  &  05 41  41.6 & -01 54  8  &  12.62  \\
 72 & &   2150 &  2160  &  05 41  41.6 & -01 54  7 &  13.36  \\
 73 & &   2050 &  1830  &  05 41  41.9 & -01 54  10  &  10.86  \\
 74$^{*}$ & &   1830 &  2235  &  05 41  42.4 & -01 54  4  &  11.72  \\
 75$^{*}$ &  &  1800 &  2435  &  05 41  42.4 & -01 53  58  &  12.22  \\
 76$^{*}$ & &   1810 &  1210  &  05 41  42.5 & -01 54  41  &  11.17  \\
 77$^{*}$ & &   1735 &  1390  &  05 41  42.6 & -01 54  34 &  14.01  \\
 78$^{*}$ & &   1715 &  1490  &  05 41  42.7 & -01 54  31  &  10.92  \\
 79$^{*}$ & &   1550 &  2590  &  05 41  43.0 & -01 53  52  &  12.32  \\
 80$^{*}$ & &   1550 &  1820  &  05 41  43.1 & -01 54  18 &  15.31  \\
 81$^{*}$ & &   1530 &  2085  &  05 41  43.1 & -01 54  10  &  14.99  \\
 82$^{*}$ & &   1520 &  1530  &  05 41  43.1 & -01 54  29  &  13.18  \\
 83$^{*}$ & &   1170 &  1220  &  05 41  44.0 & -01 54  40  &  12.64 \\
 84$^{*}$ & &   1040 &  1915  &  05 41  44.2 & -01 54  15  &  12.70  \\
 85$^{*}$ & &   840 &  1765  &  05 41  44.7 & -01 54  20  &  9.06 \\
 86$^{*}$ & &   850 &  1470  &  05 41  44.7 & -01 54  31  &  11.18 \\
 87$^{*}$ & IRS 18 &   730 &  1135  &  05 41  45.0 & -01 54  42  &  9.95 \\
 88$^{*}$ & &   690 &  1810  &  05 41  45.1 & -01 54  18  &  11.89 \\
 89$^{*}$ & &   670 &  1830  &  05 41  45.1 & -01 54  18  &  11.69 \\
 90$^{*}$ &  &  550 &  1660  &  05 41  45.4 & -01 54  23 &  7.50 \\
 91$^{*}$ &  &  475 &  1060  &  05 41  45.6 & -01 54  44  &  12.18 \\
 92$^{*}$ & NGC 2024 IRS \#2 &  420 &  1615  &  05 41  45.8 & -01 54  26 &  4.81 \\
 93$^{*}$ &  &  270 &  2050  &  05 41  46.0 & -01 54  9  &  12.42 \\
 94$^{*}$ &  &  100 &  1135  &  05 41  46.4 & -01 54  42  &  12.10 \\
\enddata


\tablenotetext{a}{Alternate name designation from the surveys of Barnes et al. (1989) and Grasdalen (1974).}
\tablenotetext{*}{Stars observed in frames without adaptive optics correction.}


\end{deluxetable}

\clearpage

\begin{deluxetable}{ccccc}
\tabletypesize{\scriptsize}
\tablecaption {Binary Stars in NGC 2024. \label{tbl-2}}
\tablewidth{0pt}
\tablehead{
\colhead{AO primary} & \colhead{AO companion}   & \colhead{ K$'$ mag (primary)$^{a}$}   &
\colhead{$\Delta$K$'$} &
\colhead{Separation ($''$)$^{b}$}  
}
\startdata
      16  &    20  &    11.49 &  1.20 & 1.87 \\ 
      17  &    18  &    10.6 &  0.5 &  0.18   \\ 
      24  &    23  &    11.6 &  0.6 &  0.42 \\  
      55  &    54  &    10.24 &   0.48 &  0.55  \\   
      64  &    63  &    13.3 &   0.1 &  0.39 \\   
      67  &    64  &    11.33 &   2.57 &  2.23  \\   
      70  &    71  &    12.19  &  0.35 &  1.85 \\   
      70  &    72  &    12.19 &  1.00 &  1.95   \\  
      88  &    89  &    11.89 &   0.20 &  0.98  \\  
 \enddata
\tablenotetext{a}{Typical uncertainty in the magnitudes of the binary 
components is $\pm$0.2 magnitudes for systems with separations less than
 0.$''$5, and $\pm$0.08 for systems with wider separations.}
\tablenotetext{b}{The uncertainty in the binary separations is $\pm$0.$''$04 
for systems closer than $\sim$0.$''$5, and $\sim$0.$''$02 for wider systems.}

\end{deluxetable}

\clearpage

\begin{figure}
\plotone{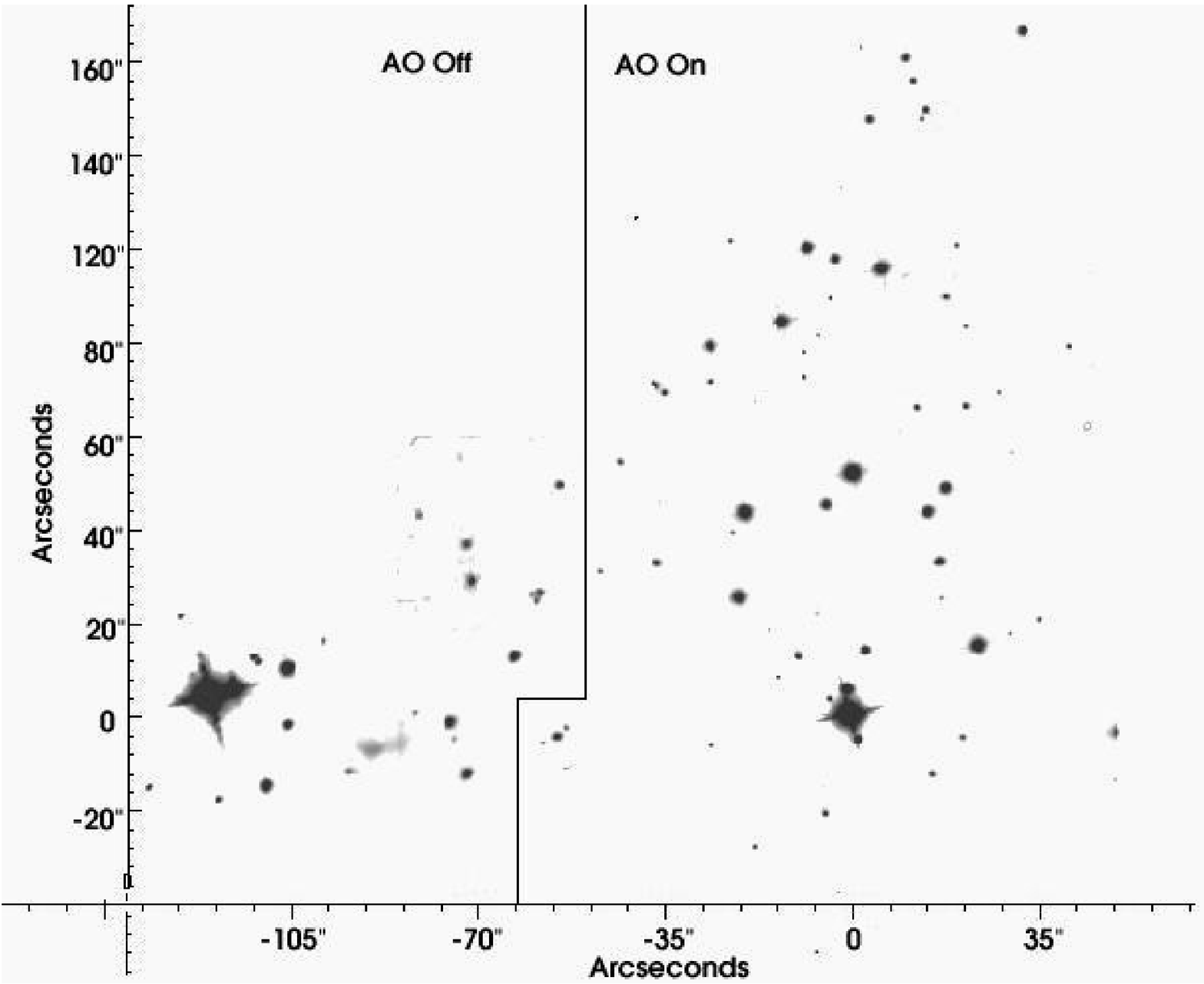}
\end{figure}

\begin{figure}
\plotone{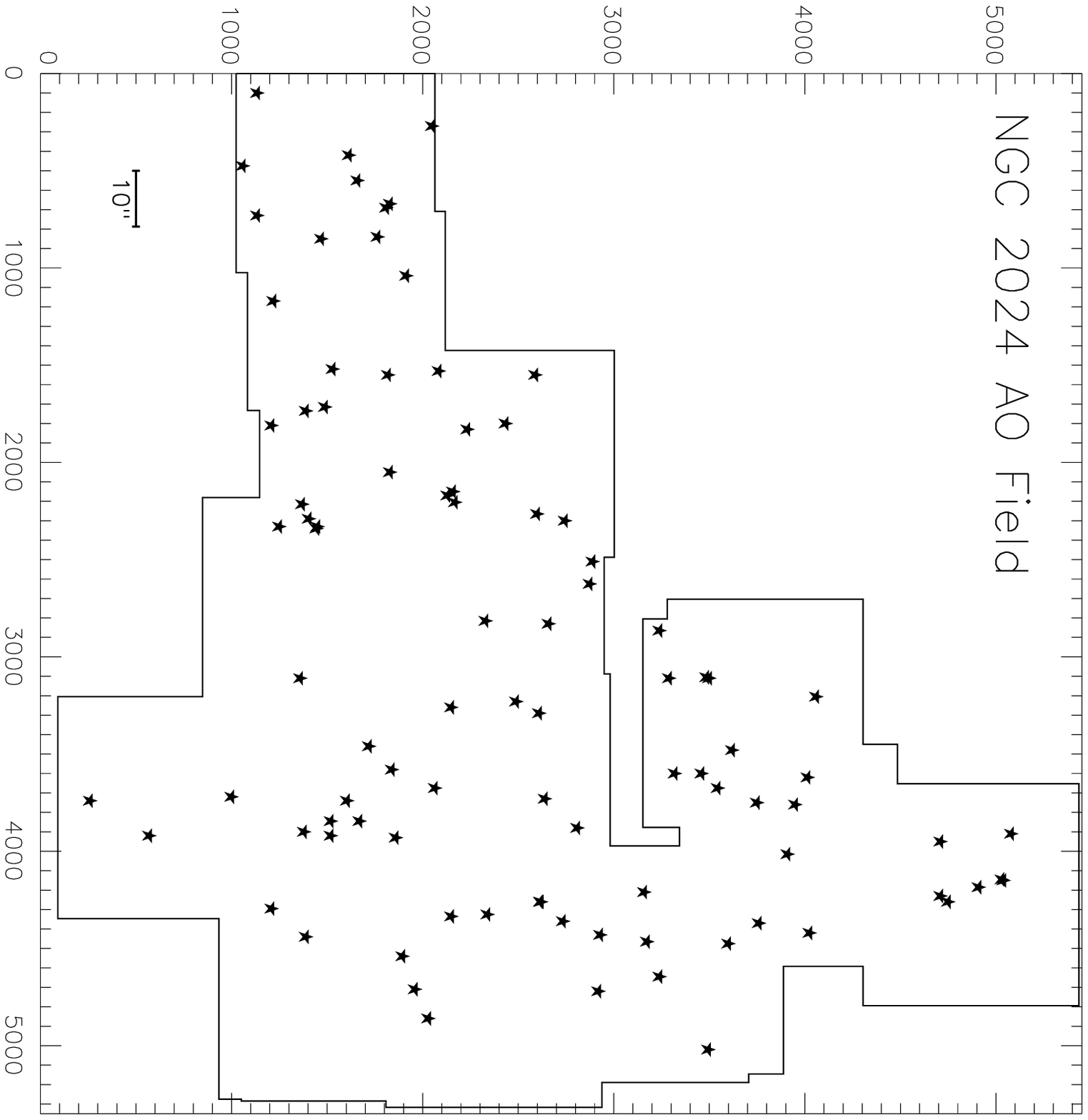}
\end{figure}

\begin{figure}
\plotone{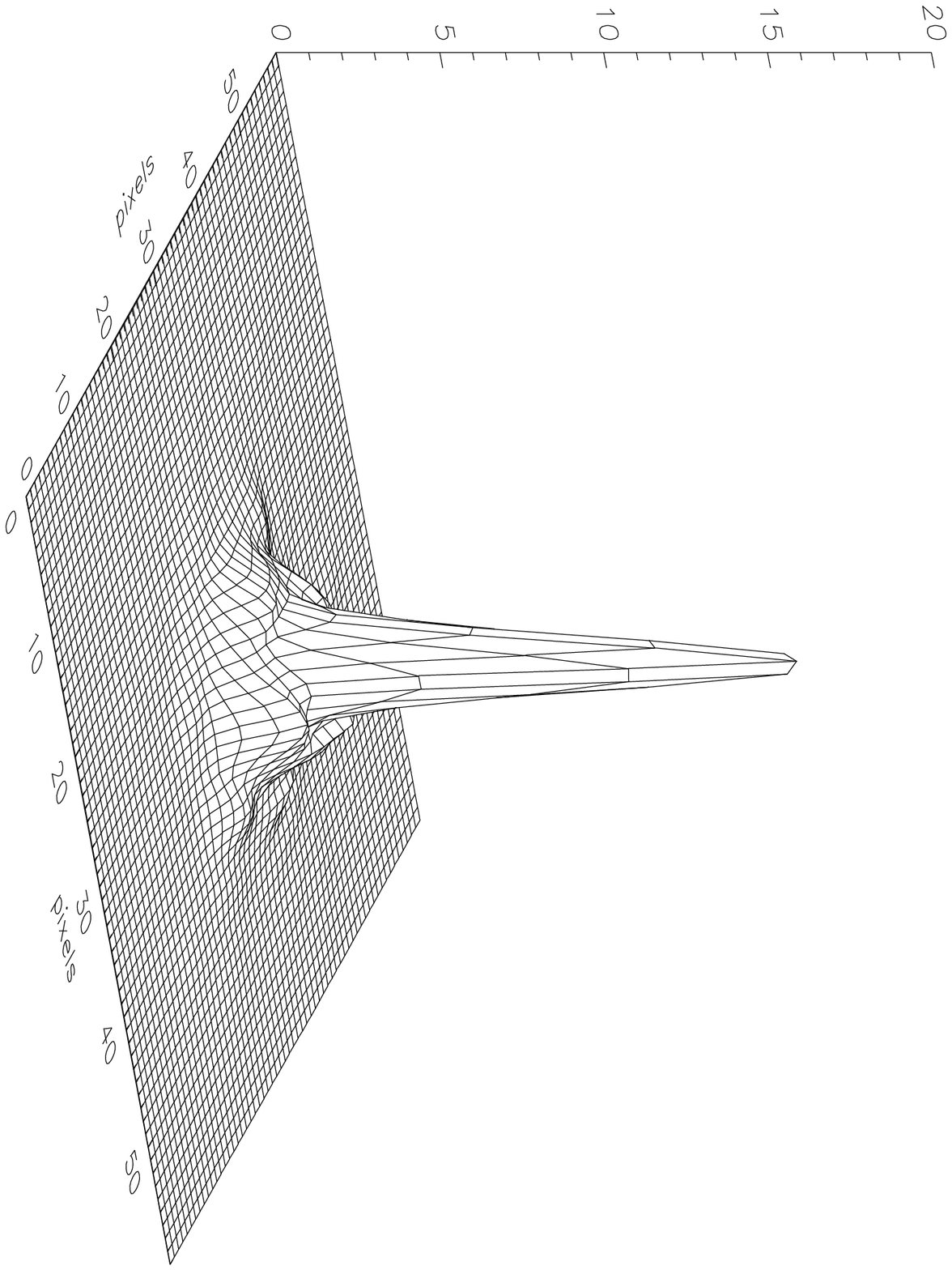}
\end{figure}

\begin{figure}
\plotone{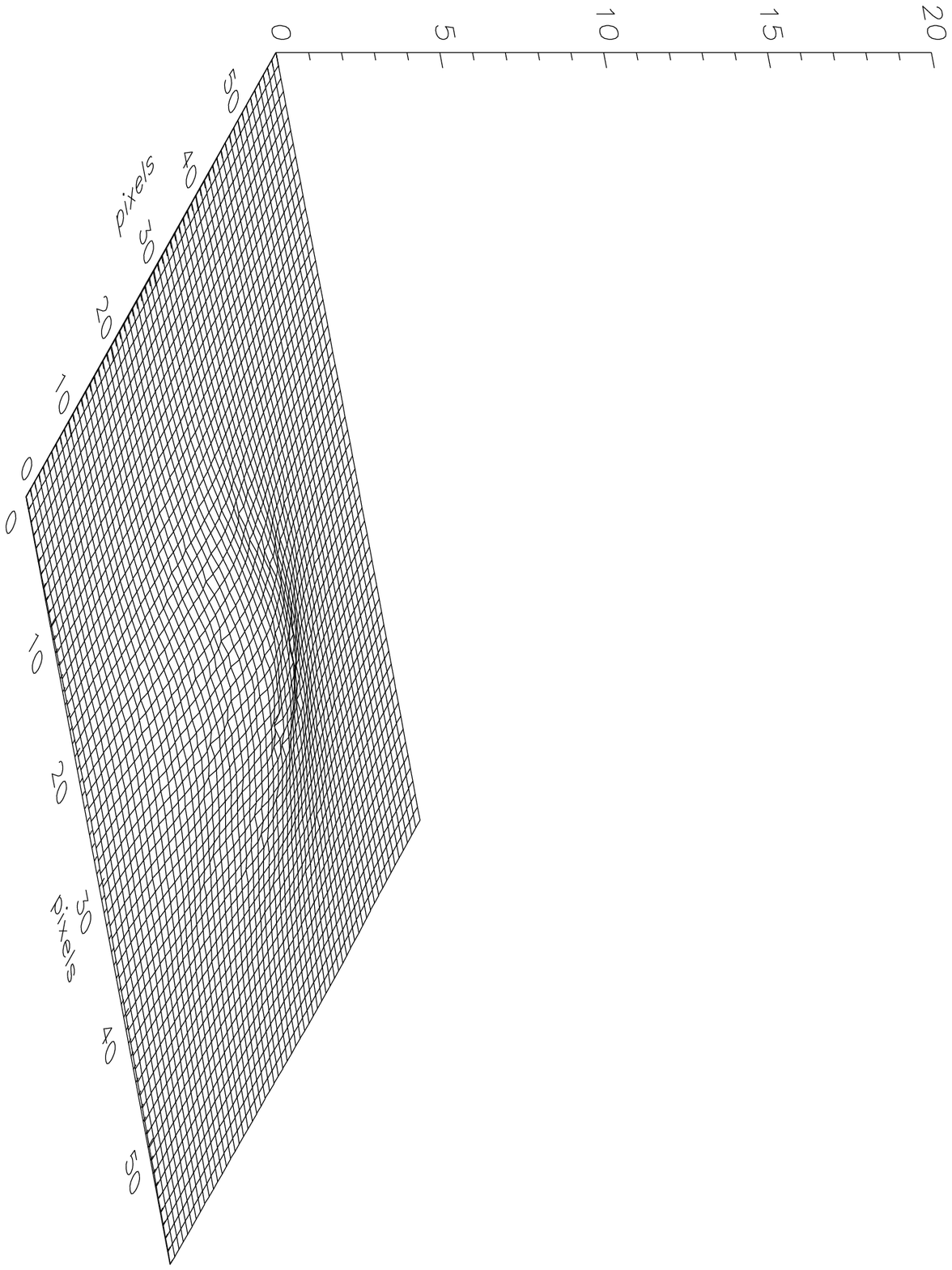}
\end{figure}

\begin{figure}
\plotone{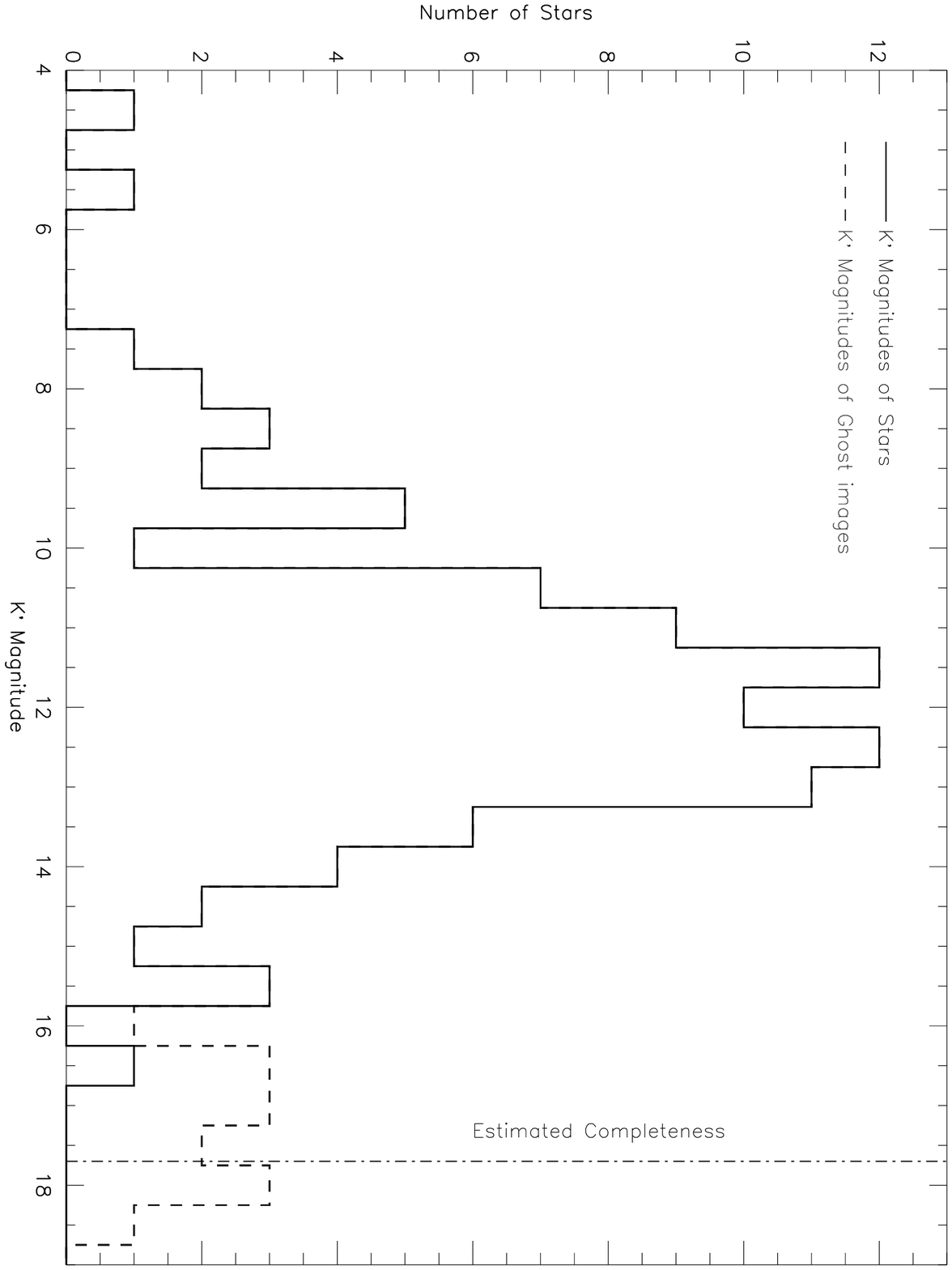}
\end{figure}

\begin{figure}
\plotone{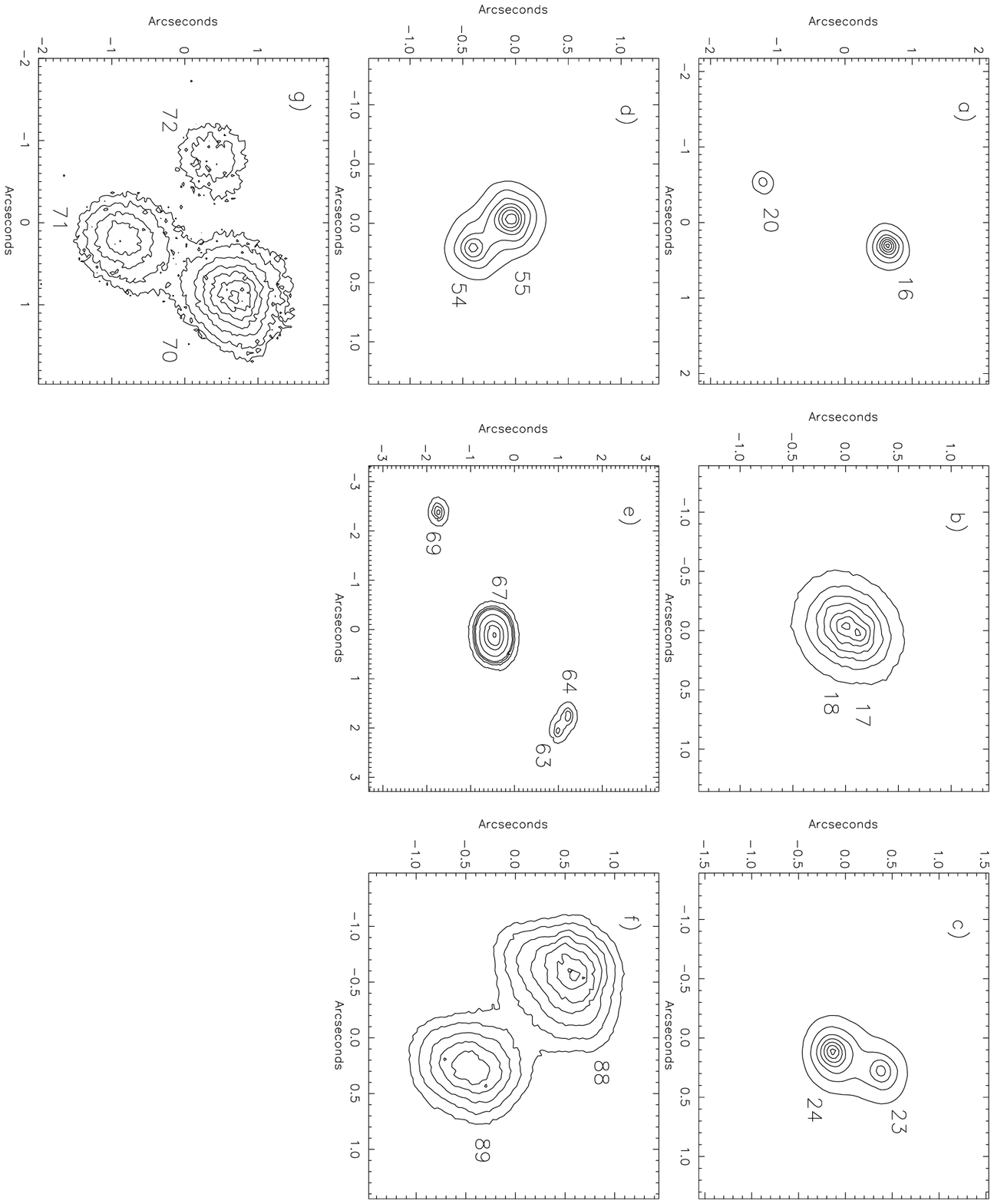}
\end{figure}

\begin{figure}
\plotone{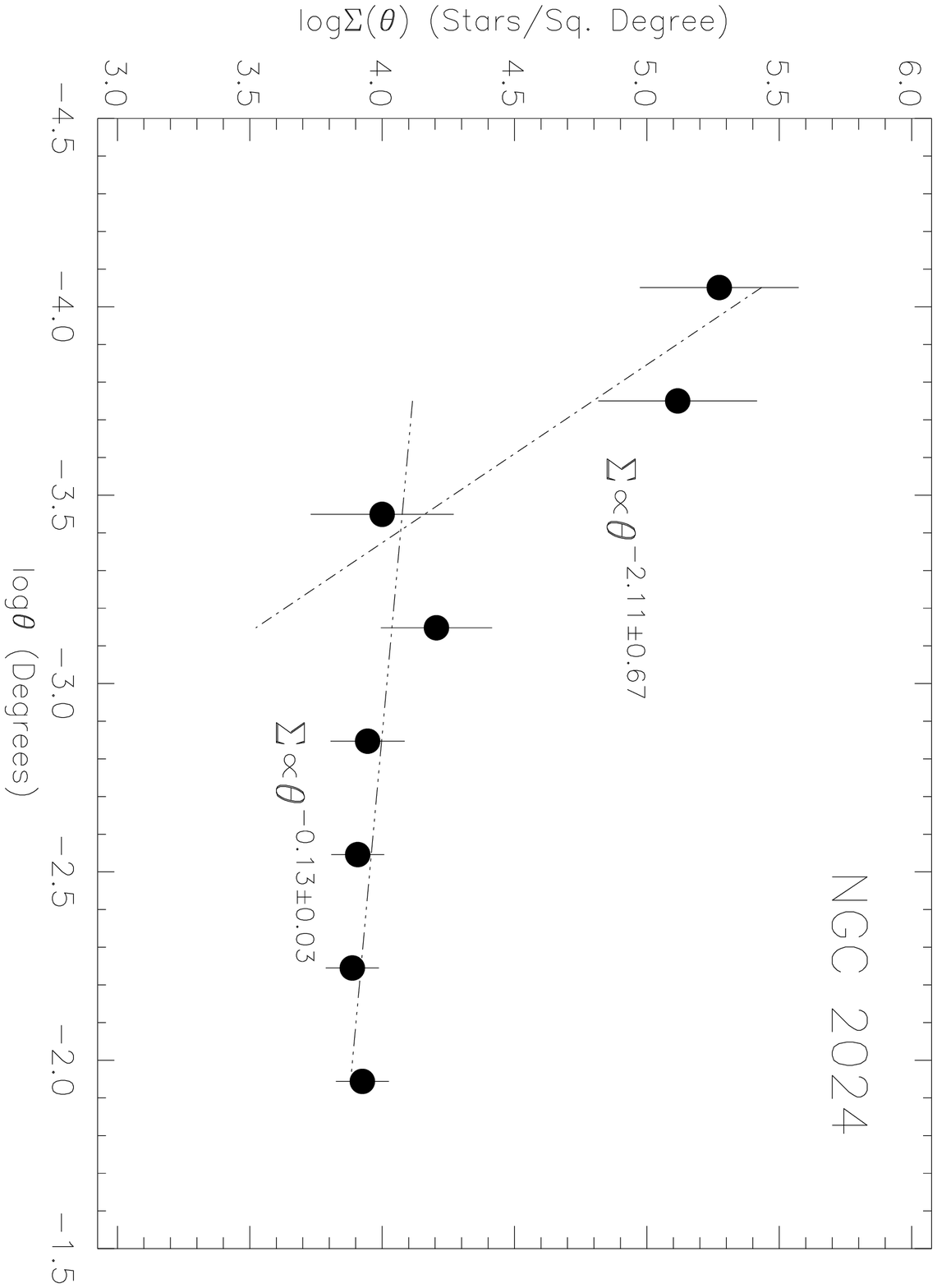}
\end{figure}

\begin{figure}
\plotone{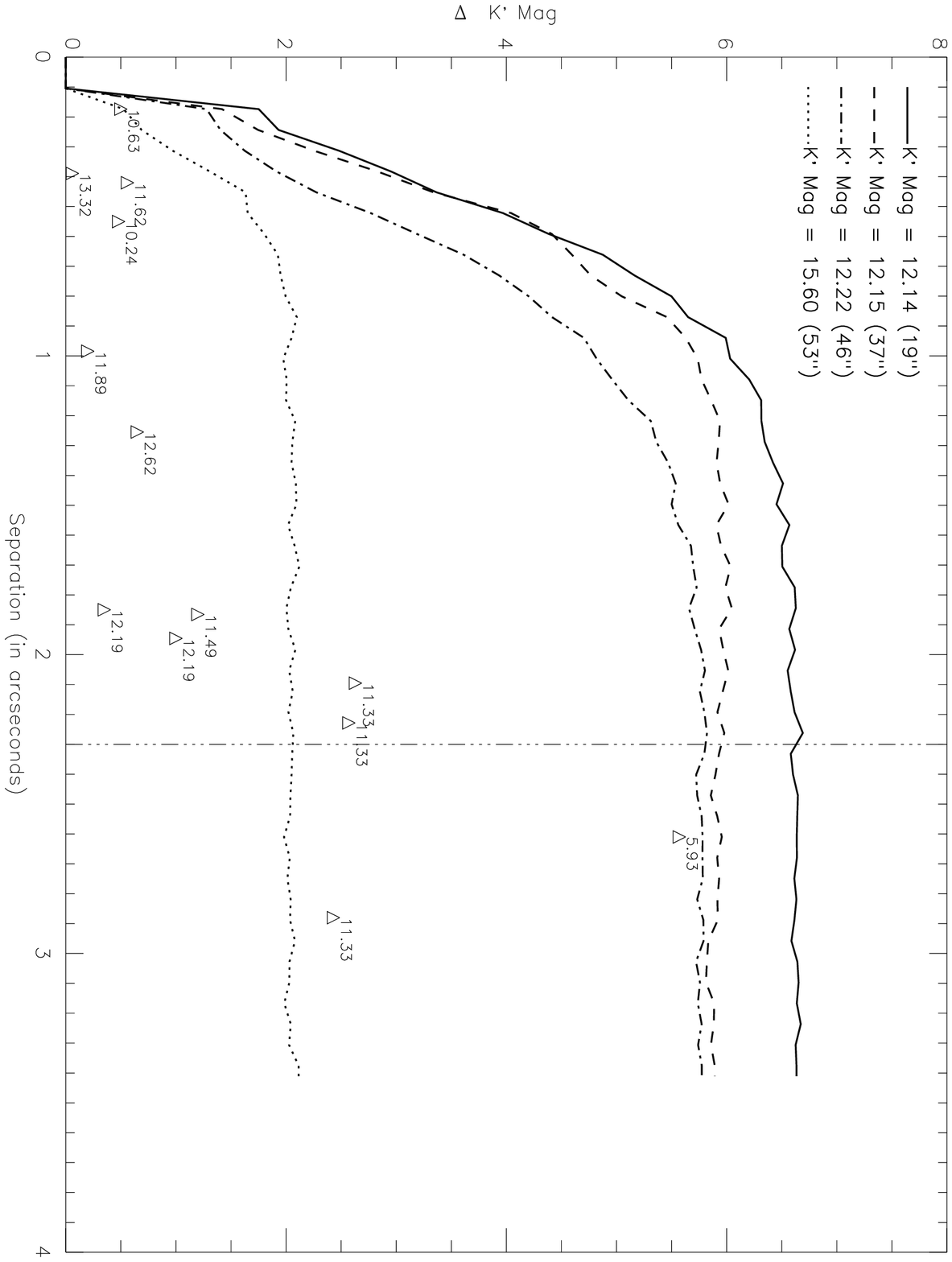}
\end{figure}

\begin{figure}
\plotone{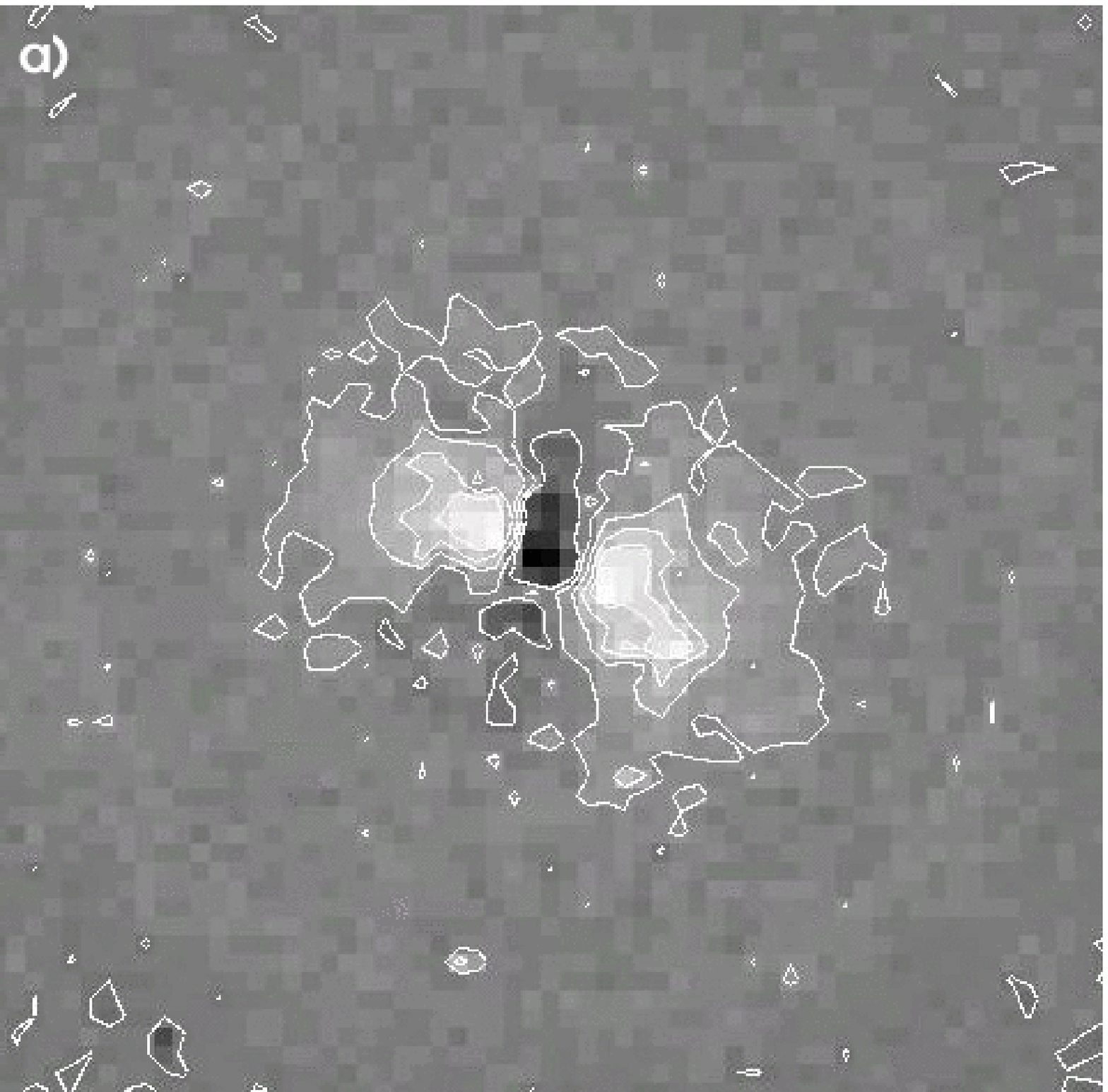}
\end{figure}

\begin{figure}
\plotone{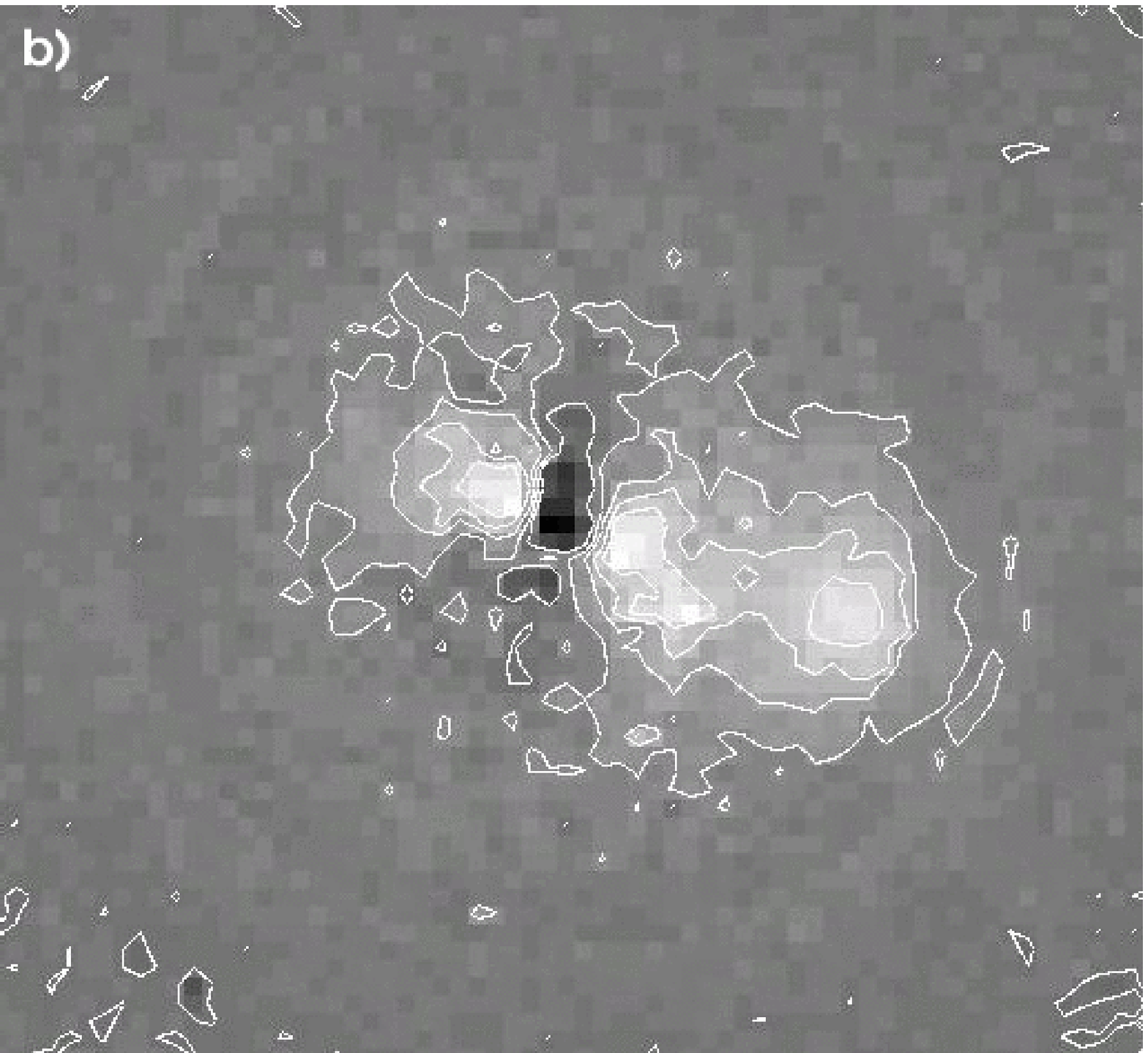}
\end{figure}

\begin{figure}
\plotone{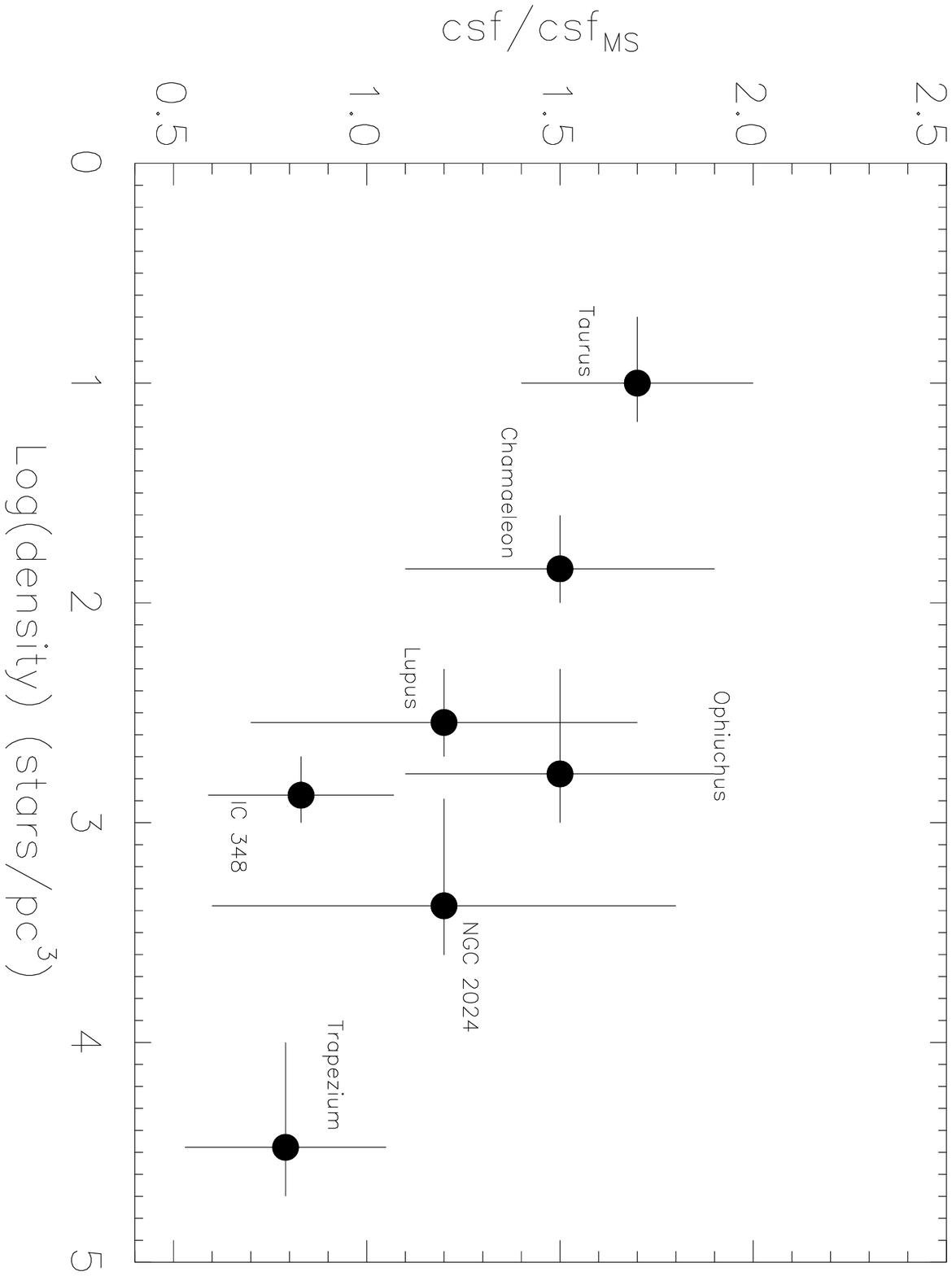}
\end{figure}

\end{document}